\documentclass[aps,article,showpacs,showkeys]{revtex4}%
\usepackage{amsmath,bm}
\usepackage{amsmath}
\usepackage{amsfonts}
\usepackage{amssymb}
\usepackage{graphicx}%
\input{psfig.sty}

\begin{document}
\title{High integer spins beyond the Fierz-Pauli Framework}
\author{S. G\'{o}mez-Avila$^{1}$, Mauro Napsuciale$^{1}$, J.A. Nieto$^{2}$, M.
Kirchbach$^{3}$}
\affiliation{$^{1}$Instituto de F\'{\i}sica, Univ. de Guanajuato \\
Lomas del Bosque 103, Fracc. Lomas del Campestre, \\
37150, Le\'{o}n, Guanajuato, M\'{e}xico. \\
$^{2}$ Fac. de Ciencias F\'{\i}sico-Matem\'{a}ticas, Univ. Aut. de Sinaloa \\
Avenida de las Am\'{e}ricas y Blvd. Universitarios, C.U. \\
80010, Culiac\'{a}n, Sinaloa, M\'{e}xico.\\
$^{3}$Instituto de F\'{\i}sica, Univ. Aut. de San Luis Potos\'{\i} \\
Av. Manuel Nava 6, San Luis Potos\'{\i}, S.L.P. 78290, M\'exico\\}
\date{\today }

\begin{abstract}
We explicitly construct in the $(\frac{1}{2},\frac{1}{2})\otimes(\frac{1}%
{2},\frac{1}{2})$ representation space the operator of the squared
Pauli-Lubanski vector and derive from it that the $\left(  -6m^{2}\right)  $
eigensubspace (spin 2 in the rest frame), with well defined parity, is pinned down
by the one sole equation, $[\epsilon_{\mu\alpha\beta\sigma}\epsilon_{\text{
\ }\nu\delta\rho}^{\mu}p^{\sigma}p^{\rho}-m^{2}g_{\alpha\beta}g_{\nu\delta
}]h^{\beta\delta}=0$. 

\end{abstract}
\maketitle

The description of higher integer spins takes its origin from the 1939 paper
by Fierz and Pauli\cite{FP} who suggested to consider spin $s$ as the highest
spin in the symmetric traceless tensor $\varphi_{\mu_{1}\mu_{2}..\mu_{s}%
}^{(s)}(x)$ belonging to the $(\frac{s}{2},\frac{s}{2})$ representation of the
Lorentz Group. This representation is reducible and contains all values of
spin from spin $0$ to $s$. All the lower spin values should be eliminated. A
result which can be accomplished by imposing the divergenless condition in
addition to the Klein-Gordon condition, i.e. the field $\varphi_{\mu_{1}%
\mu_{2}..\mu_{s}}^{(s)}(x)$ should satisfy
\begin{align}
\varphi_{\mu_{1}..\mu_{i}...\mu_{j}..\mu_{s}}^{(s)}(x)  &  =\varphi_{\mu
_{1}..\mu_{j}..\mu_{i}..\mu_{s}}^{(s)}(x),\label{FP}\\
(p^{2}-m^{2})\varphi_{\mu_{1}\mu_{2}..\mu_{s}}^{(s)}(x)  &  =0,\\
p^{\mu_{1}}\varphi_{\mu_{1}\mu_{2}..\mu_{s}}^{(s)}(x)  &  =0\,
\label{2nd_aux_cond}%
\end{align}
It was noticed in \cite{FP} that minimally coupling electromagnetism in
Eqs.(\ref{FP}) lead to immediate algebraic inconsistencies which can be
avoided by requiring that all equations involving derivatives must be
obtainable from a Lagrangian. Many attempts have been made in the past to
construct such Lagrangians \cite{Fronsdal} a procedure that becomes involved
because of the need of the use auxiliary fields in the formalism in order to
obtain Eqs.(\ref{FP}) from the Lagrangian (see Ref. \cite{FP}).

A similar formalism was also developed in \cite{FP} for the fermion case.
However, an alternative framework using symmetric spinor-vector quantities
$\psi_{\mu_{1}\mu_{2}..\mu_{s}}$ to describe spin $s=k+\frac{1}{2}$ was
formulated in \cite{RS} where a linear equation of motion is obtained from a
variational principle without the need for auxiliary fields. The corresponding
subsidiary conditions which eliminate the redundant components are obtained
from this equation. Soon after its formulation it was discovered that the
Rarita-Schwinger formalism suffer from serious inconsistencies
\cite{sudarshan,VZ1,VZ2,offshell} a problem which has been unresolved since.

In a previous work \cite{KNE} two of the authors proposed a formalism to avoid
these problems in the fermionic case. Indeed, based on the facts shown in
\cite{MC} that \ i) Dirac equation is just the projector over parity
eigensubspaces contained in $(\frac{1}{2},0)\oplus(0,\frac{1}{2})$ ii) Proca
equation is just the projector over the negative parity egensubspace contained
in $(\frac{1}{2},\frac{1}{2})$; and the fact shown in \cite{KNE} that Proca
equation can also be seen as the projector over the $-2m^{2}$ eigensubspace of
$W^{2}$ contained in $(\frac{1}{2},\frac{1}{2})$, it was concluded that\textit{
the correct equation of motion for a field should be obtained by projecting
over the corresponding eigensubspace of $W^{2}$ and parity }in a
given representation of the Lorentz Group. Guided by this principle, a new
equation of motion for a spin 3/2 particle was formulated in \cite{KNE} which
yields the appropriate subsidiary conditions and can be derived from a
variational principle without the need for auxiliary fields. Furthermore,
based on a previous study on the propagation of higher spin waves in the
framework of projectors over eigensubspaces of $W^{2}$ alone \cite{NK03}, it
was argued that this equation is free of the Velo-Zwanziger pathologies.

In this work we apply the same principle to the bosonic case. We construct
the equation of motion corresponding to the covariant projectors onto
invariant subspaces of the squared Pauli-Lubanski vector and parity.
Specifically, we explicitly construct covariant projectors in the
representation $(\frac{1}{2},\frac{1}{2})^{s}$ and derive the condition that fixes
the invariant subspace of eigenvalue $\left(  -m^{2}s(s+1)\right)  $. 
For the sake of transparency
and without any loss of generality in the following we carry out all
considerations in momentum space and we focus in the case $s=2$ .The
generalization to arbitrary integer $s$ is straightforward.

We will work with the $(\frac{1}{2},\frac{1}{2})\otimes(\frac{1}{2},\frac
{1}{2})$ representation of the Lorentz Group. This representation is reducible
and can be decomposed into irreducible representations as%
\[
(\frac{1}{2},\frac{1}{2})\otimes(\frac{1}{2},\frac{1}{2})=[(1,0)\oplus
(0,1)]\oplus(1,1)\oplus(0,0),
\]
corresponding to the direct sum of an antisymmetric second rank tensor, a
symmetric traceless second rank tensor and a scalar. It is the $-6m^{2}$
eigensubspace of $W^{2}$ (spin 2 in the rest frame) contained in the $(1,1)$
representation space (traceless symmetric tensor) which we are interested in
here. It is the goal of this paper to show that the highest spin in
$\varphi_{\mu_{1}...\mu_{s}}^{(s)}(x)$, can be pinned down by one sole
covariant equation quadratic in the momenta.

We begin with recalling that the Pauli--Lubanski (PL) vector for a given
Homogeneous Lorentz Group (HLG) representation is defined as
\begin{equation}
W_{\mu}=\frac{1}{2}\epsilon_{\mu\nu\alpha\beta}M^{\nu\alpha}P^{\beta}\,,
\label{paulu}%
\end{equation}
where $\epsilon_{0123}=1$, while $M^{\nu\alpha}$ are the corresponding
generators. This operator can be shown to satisfy the commutators
\begin{equation}
\lbrack W_{\alpha},M_{\mu\nu}]=i(g_{\alpha\mu}W_{\nu}-g_{\alpha\nu}W_{\mu
}),\qquad\lbrack W_{\alpha},P_{\mu}]=0, \label{conmrelpl}%
\end{equation}
i.e. it transforms as a four-vector under Lorentz transformations. Here
$g_{\alpha\mu}$ is a flat metric.

We first construct the generators of the HLG for the $(\frac{1}{2},\frac{1}%
{2})$ representation as
\begin{equation}
(M^{\mu\nu})_{\alpha}^{\quad\beta}=i(g_{\alpha}^{\quad\mu}g^{\nu\beta}%
-g^{\mu\beta}g_{\alpha}^{\quad\nu}).\label{generators_hh}%
\end{equation}
The Pauli-Lubanski operator in $(\frac{1}{2},\frac
{1}{2})$ space denoted by $w_{\mu}$. In substituting Eq.~(\ref{generators_hh}) into the defining
expression (\ref{paulu}) one finds
\begin{equation}
(w_{\mu})_{\alpha}^{\quad\beta}=\frac{1}{2}\epsilon_{\mu\nu\rho\sigma}%
(M^{\nu\rho})_{\alpha}^{\quad\beta}p^{\sigma}=i\epsilon_{\mu\alpha\quad\sigma
}^{\quad\ \beta}p^{\sigma}\,.\label{PL_hh}%
\end{equation}
{}From the latter equation one easily deduces $w^{2}$ as
\begin{equation}
(w^{2})_{\alpha}^{\quad\beta}=-2\left(  g_{\alpha}^{\beta}p^{2}-p_{\alpha
}p^{\beta}\right)  \,.\label{W2_hh}%
\end{equation}
This operator possesses the two different eigenvalues, $0$, and -$2p^{2}$,
respectively. States belonging to the latter eigensubspace satisfy
\begin{equation}
w^{2}A=-2m^{2}A\,.\label{hh_1}%
\end{equation}
The matrix form of $w^{2}$ implies
\begin{equation}
(w^{2}A)_{\alpha}\equiv(w^{2})_{\alpha}^{~\beta}A_{\beta}\,.\label{hh_2}%
\end{equation}
Substitution of Eq.~(\ref{W2_hh}) into Eq.~(\ref{hh_2}) under usage of
Eq.~(\ref{hh_1}) amounts to
\begin{equation}
\left(  g_{\alpha}^{\beta}p^{2}-p_{\alpha}p^{\beta}\right)  A_{\beta}%
=m^{2}A_{\alpha}\,,\label{hh_3}%
\end{equation}
or in more conventional form
\[
p^{\beta}F_{\beta\alpha}-m^{2}A_{\alpha}=0,
\]
where $F_{\beta\alpha}=p_{\beta}A_{\alpha}-p_{\alpha}A_{\beta}$. The
conclusion is that Proca equation originates directly from the frame
independent projector onto the $W^{2}$ eigensubspace that gives spin 1 at
rest. As shown in \cite{MC} , Proca operator is just the covariant
parity\ projector for this representation, hence in this space, projection
over parity eigensubspaces is equivalent to the projection over $w^{2}$ eigensubspaces.

Next we construct the Pauli-Lubanski vector for the $(\frac{1}{2},\frac{1}%
{2})\otimes(\frac{1}{2},\frac{1}{2})$ representation. The HLG generators for
the $(\frac{1}{2},\frac{1}{2})\otimes(\frac{1}{2},\frac{1}{2})$ representation
is obtained as
\begin{equation}
M_{RS}^{\mu\nu}=M_{(\frac{1}{2},\frac{1}{2})}^{\mu\nu}\otimes1_{4}%
+1_{4}\otimes M_{(\frac{1}{2},\frac{1}{2})}^{\mu\nu}\,,\label{ER_gens}%
\end{equation}
where $M_{(\frac{1}{2},\frac{1}{2})}^{\mu\nu}$, denote the generators in the
four vector space, while $1_{4}$ is the four dimensional unit matrix. The
Pauli-Lubanski vector for this representation is then calculated as
\begin{equation}
(W_{\mu})_{\alpha\beta ab}=(w_{\mu})_{\alpha\beta}\ g_{ab}+g_{\alpha\beta
}\ (w_{\mu})_{ab},\label{W_RS}%
\end{equation}
where we consider that both Greek and Roman indices are Lorentz ones and we
make a distinction only to keep track of quantities coming from the different
$(\frac{1}{2},\frac{1}{2})$ subspaces. The squared operator is calculated as
\begin{equation}
(W^{2})_{\alpha\beta ab}=(w^{2})_{\alpha\beta}\text{ }g_{ab}+g_{\alpha\beta
}\text{ }(w^{2})_{ab}+2[(w_{\mu})_{\alpha\gamma}g_{ac}][(w^{\mu}%
)_{cb}g_{\gamma\beta}].\label{proj}%
\end{equation}
Next we choose the appropriate $W^{2}$ and parity eigensubspace by restricting
the state field $h_{\alpha a}$ to satisfy%
\[
(w^{2})_{\alpha\beta}\text{ }g_{ab}\text{ }h^{\beta b}=-2m^{2}h_{\alpha
a}\qquad g_{\alpha\beta}\text{ }(w^{2})_{ab}\text{ }h^{\beta b}=-2m^{2}%
h_{\alpha a},
\]
i.e to be composed as the direct product of fields in $(\frac{1}{2},\frac
{1}{2})$ belonging to the $-2m^{2}$eigensubspace of $w^{2}$. Finally, we
impose that $h^{\beta b}$ belongs to the $-6m^{2}$ eigensubspace in the full
$(\frac{1}{2},\frac{1}{2})\otimes(\frac{1}{2},\frac{1}{2})$ space, thus we
get
\begin{equation}
-6m^{2}h_{\alpha a}=-2m^{2}h_{\alpha a}-2m^{2}h_{\alpha a}+2[(w_{\mu}%
)_{\alpha\gamma}g_{ac}][(w^{\mu})_{cb}g_{\gamma\beta}]h^{\beta b}.\label{eom}%
\end{equation}
Using the explicit form in Eq.(\ref{PL_hh}) we obtain the equation of motion
for a massive spin 2 particle with well defined parity as
\begin{equation}
\lbrack\epsilon_{\mu\alpha\beta\sigma}\epsilon_{\ abc}^{\mu}p^{\sigma}%
p^{c}-m^{2}g_{\alpha\beta}g_{ab}]\text{\ }h^{\beta b}=0.\label{diamond}%
\end{equation}
Equation (\ref{diamond}) is our prime result. It defines uniquely the
$-6m^{2}$ eigensubspace of $W^{2}$ with well defined parity in the $(\frac
{1}{2},\frac{1}{2})\otimes(\frac{1}{2},\frac{1}{2})$ representation space. It
follows directly from the symmetries of space-time and is unambiguous.
Furthermore, in general second order equations containing factors of  $p^{\mu
}p^{\alpha}$ are ambiguous with respect to the order of these factors under
gauging (see e.g. \cite{Novello}). This is not the case for our equation since
we keep track of the $w^{2}$ sectors where the $p$ factors come from.

In principle $h^{\beta b}$ contains more components than necessary to describe
a spin 2 particle. However, from Eq.(\ref{diamond}) we obtain subsidiary
conditions which eliminate the redundant components. Indeed, the first
property of this field which is obvious from this equation is its symmetry
under the exchange of Lorentz indices%
\begin{equation}
h_{\alpha\beta}=h_{\beta\alpha}. 
\label{symmetry}%
\end{equation}
The second set of conditions are obtained contracting our Eq.(\ref{diamond})
with $p^{\alpha}$, due to the anti-symmetry of the Levi-Civita tensor one
easily finds
\begin{equation}
p_{\beta}h^{\beta a}=0\,. 
\label{divergenless}%
\end{equation}
Finally going to the rest frame we can easily convince ourselves that in this
frame
\begin{equation}
h_{\alpha}^{~\alpha}=0, 
\label{cero}%
\end{equation}
and since this is a scalar quantity this condition is fulfilled in every
frame. Eqs.(\ref{symmetry},\ref{divergenless},\ref{cero}) eliminate 11 of the 16
components contained in $h_{\alpha\beta}$ and we are left with only $5$
degrees of freedom which are the same number of d.o.f. for a massive spin 2 particle.

It is worth remarking that in the massless case this equation coincides with
the one satisfied by the graviton field in sourceless linearized gravity which
describes a free graviton in a flat Minkowski space \cite{Wheeler} (see also
\cite{nieto}). The propagation of spin 2 waves under minimal coupling in our
formalism is presently under investigation. In this concern, our work may be
also useful to clarify some aspects about the relation between the mass of the
graviton and the cosmological constant $\Lambda$ which recently has been
subject of some interest \cite{Novello,Deser} in connection with
causality for the propagation of a graviton in an electromagnetic background.
In this direction, it is worth mentioning that it has been shown \cite{nieto2}
that linearized gravity with cosmological constant admits an $S$ duality
prescription $\Lambda\rightarrow\frac{1}{\Lambda}$ or a strong coupling limit
\cite{Hull} via the duality $l_{p}\rightarrow\frac{1}{l_{p}}$. It may be
interesting for further research to see if our present study is useful in this direction.

\vspace{0.5cm} Work supported by Consejo Nacional de Ciencia y Tecnologia
(CONACYT) Mexico under projects 37234-E and C01-39820 .


\begin{thebibliography}{99}                                                                                               %


\bibitem {FP}M. Fierz, W. Pauli, Proc. Roy. Soc. (London) \textbf{A173}, 211 (1939).

\bibitem {Fronsdal}C. Fronsdal, Nuovo Cimento Suppl. \textbf{9 }, 416 (1958);
J.Chang, Phys. Rev. \textbf{161}.1308 (1967); L. P. S. Singh, C. R. Hagen,
Phys. Rev. \textbf{D9}, 898 (1974).

\bibitem {RS}W.\ Rarita, J.\ Schwinger, Phys.\ Rev.\ 60, 61 (1941).

\bibitem {sudarshan}
K.~Johnson, E.~C.~Sudarshan,
Annals of Physics \ \textbf{13}, 126 (1961).


\bibitem {VZ1}G.~Velo, D.~Zwanziger,
Phys.\ Rev.\ \textbf{186}, 1337 (1969).

\bibitem {VZ2}G. Velo, D. Zwanziger,
Phys. Rev. \textbf{188}, 2218 (1969).

\bibitem {offshell}
L.~M.~Nath, B.~Etemadi, J.~D.~Kimel,
Phys.\ Rev.\ D \textbf{3}, 2153 (1971);
R.\ Davidson, N.\ C.\ Mukhopadhyay, R.\ S.\ Wittman, Phys.\ Rev.\ \textbf{D}
43, 71 (1991).

\bibitem {KNE}
M.~Kirchbach and M.~Napsuciale, ``High spins beyond Rarita-Schwinger
framework,'' arXiv:hep-ph/0407179.


\bibitem {MC}
M.~Napsuciale and C.~A.~Vaquera-Araujo, ``Equations of motion as projectors
and the gyromagnetic factor g(s) = 1/s from first principles,''
arXiv:hep-ph/0310106.


\bibitem {NK03}M.\ Napsuciale, M.\ Kirchbach, \textit{Avoiding superluminal
propagation of higher spin waves via projectors onto $W^{2}$ invariant
subspaces\/}, \texttt{E-Print ArXiv: hep-ph/0311055}. Accepted for publication
in Journal of Mathematical Physics.

\bibitem {Wheeler}Gravitation, C.W. Misner, K. S. Thorne, J.A.
Wheeler,\textit{W. H. Freeman and Co. Ed}. 1970, page.436 Eq.(18.5).

\bibitem {nieto}J. A. Nieto and O. Obregon, Phys. Lett. \textbf{A 175}, 11 (1993).

\bibitem {Novello}M. Novello and R.P. Neves, Class. Quant. Grav. \textbf{20},
L67 (2003); M. Novello, S. E. Perez Bergliaffa, R. P. Neves, "Replay
acausality of massive charged spin-2 field";\ gr-qc/0304041.

\bibitem {Deser}S. Deser and A. Waldron, "Acausality of massive charged spin-2
field", hep-th/0304050.

\bibitem {nieto2}J. A. Nieto, Phys. Lett. \textbf{A 262}, 274 (1999); hep-th/9910049.

\bibitem {Hull}C. M. Hull, Nucl. Phys. \textbf{B583}, 237 (2000); hep-th/0004195.
\end{thebibliography}
\end{document}